\documentclass[%
 reprint,
 amsmath,amssymb,
 aps,
]{revtex4-2}

\usepackage{graphicx}
\usepackage{dcolumn}
\usepackage{bm}
\usepackage[colorlinks=true, allcolors=blue]{hyperref}

\newcommand{\dd}{\mathrm{d}}
\newcommand{\ii}{\mathrm{i}}

\newcommand{\GF}{G_{\mathrm{F}}}

\begin{document}

\title{Fast neutrino-flavor swap in high-energy astrophysical environments}

\author{Masamichi Zaizen}
\email{zaizen@heap.phys.waseda.ac.jp}
\affiliation{Faculty of Science and Engineering, Waseda University, Tokyo 169-8555, Japan}

\author{Hiroki Nagakura}
\affiliation{Division of Science, National Astronomical Observatory of Japan, 2-21-1 Osawa, Mitaka, Tokyo 181-8588, Japan}

\date{\today}

\begin{abstract}
We assert that non-linear features of fast neutrino-flavor conversion (FFC) can be qualitatively different between core-collapse supernovae (CCSNe) and binary neutron star mergers (BNSMs).
This argument arises from recent global FFC simulations in BNSM, in which fast flavor swap (FFS) emerges in very narrow spatial regions, whereas neutrinos in CCSN tend to evolve towards flavor equipartition.
In this paper, we provide the physical mechanism of FFS based on a colliding neutrino beam model.
Neutrinos/antineutrinos can undergo FFS when they propagate in ambient neutrino gas that propagates in the opposite direction and also has the opposite sign of ELN-XLN, where ELN and XLN denote electron- and heavy-leptonic neutrino number, respectively.
Such environments can be naturally realized in BNSMs, whereas they are unlikely in CCSNe unless the neutrino sphere is strongly deformed aspherically.
Our study exhibits the diversity of non-linear dynamics of FFC.
\end{abstract}

\maketitle


\section{Introduction}
Neutrino flavor evolution plays important roles in the dynamics of core-collapse supernovae (CCSNe) and binary neutron star mergers (BNSMs).
During the last decade, their numerical models with classical neutrino transport schemes have cultivated our understandings of complex physical processes including weak interactions, gravity, and equation of states \cite{Langanke:2003,Janka:2012a,Janka:2017,Burrows:2021,Sumiyoshi:2023}.
Meanwhile, recent progress in neutrino quantum kinetics has indicated that flavor conversion driven by neutrino self-interactions can bring significant change in the neutrino radiation field \cite{Nagakura:2023,Shalgar:2023,Xiong:2023,Nagakura:2023b}.

Fast neutrino-flavor conversion (FFC), which is induced by a zero crossing in ELN (electron neutrino-lepton number)-XLN (heavy-leptonic one) angular distributions, has particularly received significant attention.
This phenomenon induces {\it pairwise} flavor conversion, $\nu_e\bar{\nu}_e\leftrightarrow\nu_x\bar{\nu}_x$ ($\nu_e$ and $\bar{\nu}_e$ represent electron-type neutrinos and their anti-partners, respectively;
$x$ denotes heavy-leptonic flavors, $\mu$ or $\tau$), conserving the net flavor distribution.
Linear stability analyses of flavor instabilities have been carried out for CCSN/BNSM models, and these studies have revealed that FFC occurs in both CCSNe \cite{Abbar:2019,DelfanAzari:2020,Nagakura:2019,Abbar:2020,Glas:2020,Morinaga:2020,Nagakura:2021b,Harada:2022,Akaho:2023} and BNSMs \cite{Wu:2017,George:2020,Li:2021,Richers:2022a,Just:2022,Grohs:2023,Froustey:2023}.
It should be noted, however, that we still have little knowledge concerning how the neutrino radiation field is actually changed after FFC grows substantially.
This indicates that the impacts of FFC on CCSN/BNSM remain a matter of debate \cite{Nagakura:2023,Fujimoto:2023,Ehring:2023,Ehring:2023a,Nagakura:2023d}.

It has also been suggested that flavor mixing occurring in their environments is not only FFC but also collision-induced one, so-called collisional flavor instability (CFI) \cite{Johns:2023a}, and matter-neutrino resonance (MNR) \cite{Malkus:2014}.
The former is driven by the disparity in the neutrino reaction rates with matter between neutrinos and antineutrinos.
The possibility in BNSMs \cite{Xiong:2023c} and CCSNe \cite{Liu:2023d,Akaho:2024a} has been discussed based on the linear stability analysis, and the nonlinear behaviors have been also investigated \cite{Lin:2023,Xiong:2023,Liu:2023,Kato:2023a}.
These studies showed that CFI can occur in the deeper neutrino-opaque region than FFC, indicating that they can also be a potential game-changer for BNSMs and CCSNe.

It should be mentioned that the latter (MNR) can occur in BNSMs \cite{Malkus:2014,Wu:2016a,Zhu:2016,Vaananen:2016,Frensel:2017,Tian:2017,Shalgar:2018,Vlasenko:2018,Padilla-Gay:2024}, but it is unlikely in CCSN environments.
This is because the electron-lepton number is always positive in CCSNe, which suppresses the MNR.
On the other hand, the number density of $\bar{\nu}_e$ can dominate over $\nu_e$ in some regions, which yields negative strength of neutrino self-interaction potential.
This can cancel the positive matter potential, which potentially induce MNR.
Although it is an intriguing question how these flavor conversion impact on fluid dynamics and nucleosynthesis in BNSMs, the detailed studies are postponed to another paper and we focus only on FFC in the present study.

CCSNe and BNSMs are attractive sites exhibiting rich flavor conversion phenomena, and the identification of the asymptotic states helps us to understand the impact on them.
Asymptotic states of FFCs are dictated by the interplay between neutrino advection and ELN-XLN angular distributions, where ELN and XLN represent $\nu_e - \bar{\nu}_e$ and $\nu_x - \bar{\nu}_x$, respectively.
Previous local FFC simulations suggested that ELN becomes nearly equal to XLN in some angular regions, i.e., flavor equipartition so that ELN-XLN angular crossings disappear \cite{Wu:2021,Bhattacharyya:2021,Bhattacharyya:2022,Richers:2022,Zaizen:2023}.
This can be understood analytically \cite{Zaizen:2023,Zaizen:2023a,Xiong:2023b}, and large-scale FFC simulations in spherical symmetry also support the argument \cite{Nagakura:2022a,Nagakura:2023a}.
The trend holds even in cases with including neutrino-matter interactions under realistic CCSN fluid profiles \cite{Nagakura:2023,Nagakura:2023d}.

\begin{figure*}[t]
    \centering
    \includegraphics[width=1.\linewidth]{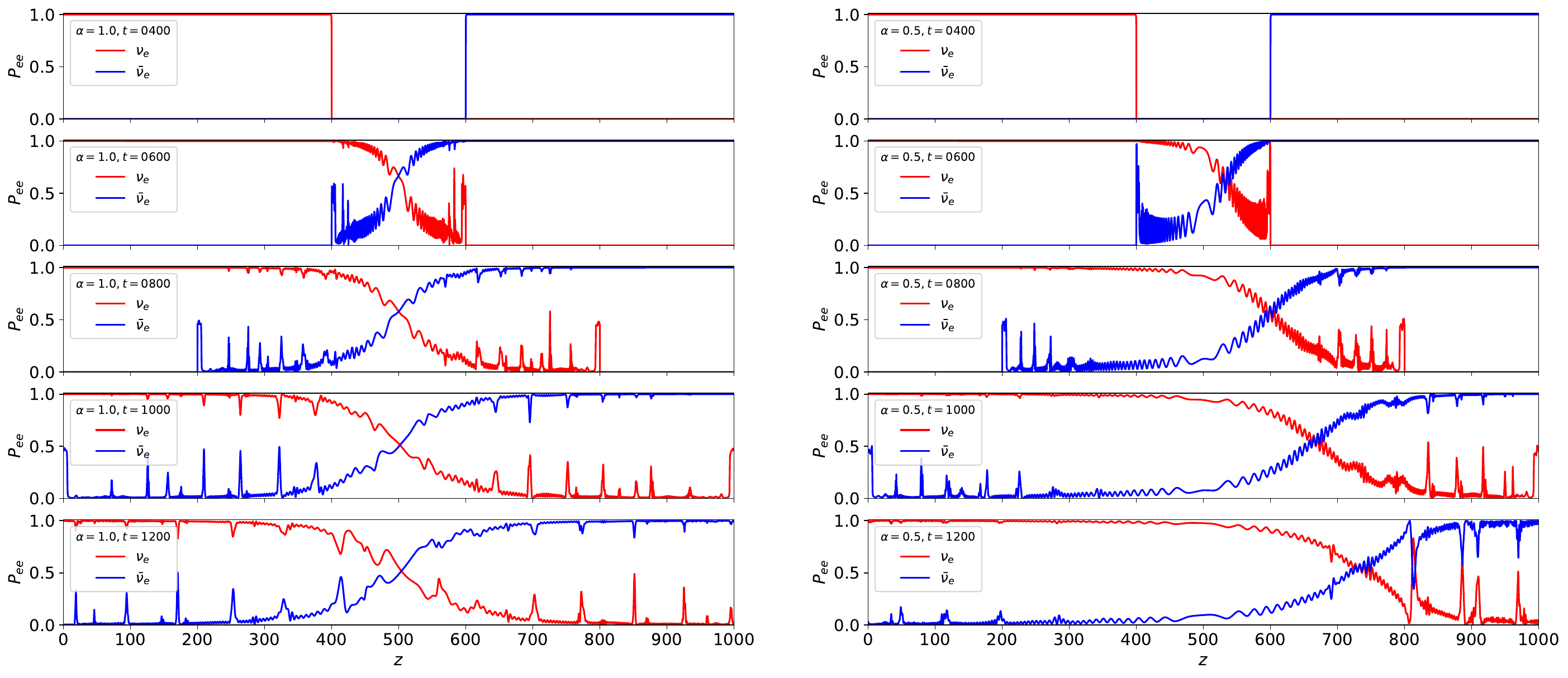}
    \caption{Time evolution of survival probability for $\nu_e$ (red) and $\bar{\nu}_e$ (blue) in the two-beam model.
    Left panels show the case of $\alpha=1$, symmetric case, and right ones the case of $\alpha=0.5$.
    Transition layer moves toward positive clearly for $\alpha=0.5$.}
    \label{fig:surv_time}
\end{figure*}

However, our large-scale FFC simulations in BNSM showed that the nonlinear evolution of FFC is qualitatively different from those in CCSNe \cite{Nagakura:2023b}.
Instead of balancing between ELN and XLN, FFC can proceed beyond the flavor equilibrium states, and the neutrinos eventually achieve flavor swap, i.e., complete interchange between different flavors.
However, the physical mechanism of fast neutrino-flavor swap (FFS) remains entirely unknown and even unexplored yet.

In this paper, we provide the principle mechanism of FFS with performing local FFC simulations.
As shall be discussed below, FFS can commonly occur in BNSM, but it seems unlikely in CCSN unless neutrino radiation fields become strongly aspherical.
The present study contributes to a comprehensive understanding of asymptotic states of FFC in CCSNe and BNSMs.

\section{Colliding neutrino beam model}\label{Sec.II}
We show that essential features of FFS can be demonstrated by a simple model, namely {\it colliding neutrino beam model}.
This is a model that neutrino beams emitted at opposite boundaries collide each other.
It should be mentioned that this model is different from canonical "beam" model which has been well studied in the literature \cite{Duan:2015a,Chakraborty:2016a,Capozzi:2017,Chakraborty:2020,Hansen:2022}.
In the canonical one, neutrino gas is assumed to be homogeneous or in the linear regime, whereas neutrino advection plays a pivotal role in our colliding beam model.
Below, we present its numerical simulations, while our numerical setup is rather simple to keep our discussions as easy as possible.

The basic equation is given by quantum kinetic equation (QKE),
\begin{equation}
    (\partial_t+v_z\partial_z)\rho(t,z,v_z) = -\ii \left[H_{\nu\nu},\rho(t,z,v_z)\right];
    \label{eq:QKE}
\end{equation}
where $\rho$, $t$, and $z$ denote the density matrix of neutrinos, time, and space, respectively.
We work in two-flavor framework and in one-dimensional ($z$ direction) transport.
We assume axial symmetry in momentum space ($v_z$ specifies the angular point in neutrino momentum space) and also ignore the energy-dependent term such as vacuum oscillation and neutrino-matter interactions, so that we consider only energy-integrated QKEs.
In this case, wave functions for anti-neutrinos can be collectively treated with those for neutrinos as negative occupations, $\bar{\rho}(E) \equiv -\rho(-E)$.
The Hamiltonian of neutrino self-interactions is recast into
\begin{equation}
    H_{\nu\nu} = \mu\int^{+1}_{-1}\dd v_z^{\prime} (1-v_z v_z^{\prime})G_{v_z^{\prime}}\rho_{v_z^{\prime}},
    \label{eq:Hvv}
\end{equation}
where $\mu=\sqrt{2}\GF n_{\nu_e}$ and $G_{v_z}$ is an ELN-XLN angular distribution;
\begin{equation}
    G_{v} = \frac{1}{n_{\nu_e}}\int\frac{E^2\dd E}{(2\pi)^2}\left[\left(f_{\nu_e}-f_{\nu_x}\right)-\left(f_{\bar{\nu}_e}-f_{\bar{\nu}_x}\right)\right].
\end{equation}
In the expression, $n_{\nu_{\eta}}$ and $f_{\nu_\eta}$ denote number density and distribution function of $\eta$-flavor neutrinos, respectively.
Hereafter, we measure the time and space in the unit of $\mu^{-1}$ because the self-interaction potential $\mu$ is a unique dimensional quantity in the energy-integrated QKE.
We cover a spatial domain of $L_z=1000$ by a uniform spatial grid points with $N_z=10000$ in our simulations.

As the simplest case, we inject neutrinos from each boundary as
\begin{equation}
    G_{v_z} = 1 \hspace{2mm} (\mathrm{for}\, v_z=1)
    \label{eq:Twobeam_vplus}
\end{equation}
at $z=0$ and
\begin{equation}
    G_{v_z} = -\alpha \hspace{2mm} (\mathrm{for}\, v_z=-1)
    \label{eq:Twobeam_vmin}
\end{equation}
at $z=L_z$, where $\alpha (>0)$ denotes a flavor asymmetry of ELN-XLN beams between those emitted from the two boundaries.
Note that $\bar{\nu}$ needs to be no longer distinguished from $\nu$ in the context of FFC.
We can consider only ELN-XLN not each species, so the number density in one beam can be normalized by that in the other.
During the simulation, we keep them constant in time at each boundary, while we put a small perturbation of $\sim 10^{-6}$ in the off-diagonal component of the density matrix only at $t=0$.
The negative sign of $G_{v_z}$ for $v_z=-1$ guarantees that ELN-XLN angular crossings appear when the two beams encounter each other, marking the onset of FFC.
Throughout the paper, we inject only $\nu_e$ at $z=0$ and $\bar{\nu}_e$ at $z=L_z$ to simplify the discussion (but without any loss of generality).

Left panels in Fig.\,\ref{fig:surv_time} show the time evolution of the survival probability of $\nu_e$ and $\bar{\nu}_e$ for $\alpha=1$, i.e., symmetric case.
As shown in the second panel from the top, FFC develops at the center of the spatial domain ($z \sim 500$) and creates a transition layer of flavor conversion there.
The flavor conversions do not stop at a flavor equilibrium state, but rather, they achieve FFS ($P_{ee} \sim 0$) eventually.
The neutrinos injected at $t>0$ also undergo FFS (see from the third to fifth panels), while the transition layer is less evolved with time.
In fact, $P_{ee}$ keeps $\sim 0.5$ at $z=500$ after the two beams are colliding.
This layer corresponds to ELN-XLN Zero Surface (EXZS), which was also observed in global FFC simulations in BNSM environment \cite{Nagakura:2023b}.
This exhibits that the colliding neutrino beam model demonstrates essentially the same phenomenon as in FFC of BNSM environments.
Before we move on to the detailed discussion of the mechanism of FFS, we present some important properties of EXZS.

\begin{figure}[t]
    \centering
    \includegraphics[width=1.\linewidth]{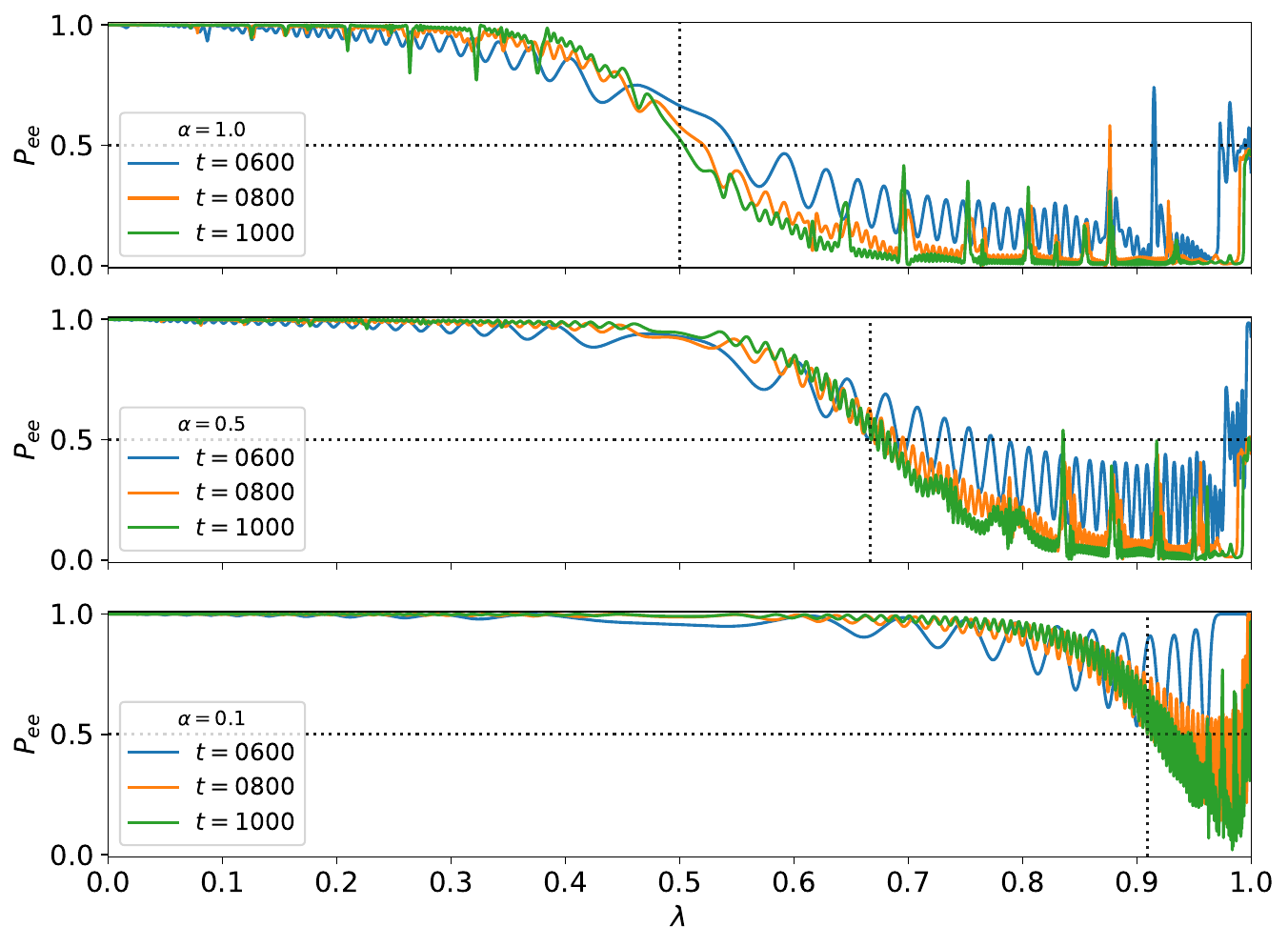}
    \caption{Time evolution of survival probability in the domain scaled by the interval between the leading parts of injected neutrinos for $\alpha=1$ (top), $0.5$ (middle), and $0.1$ (bottom).
    Dotted vertical line is $\lambda_{\mathrm{EXZS}}$.
    }
    \label{fig:scale_plot}
\end{figure}

In the case of $\alpha=1$, the transition layer including EXZS stagnates at $z \sim 500$.
This can be naturally expected, since $\nu_e$ and $\bar{\nu}_e$ beams emitted from opposite boundaries are completely symmetric.
As shown in the case with $\alpha=0.5$ (see right panels in Fig.\,\ref{fig:surv_time}), however, the layer is no longer stationary but moves toward the positive $z$-direction with time.
This is because the total number of neutrinos experiencing flavor conversions needs to be the same between $\nu_e$ and $\bar{\nu}_e$ in FFC.
In the case of $\alpha=0.5$, the $\bar{\nu}_e$ beam (emitted from $z=L_z$) is weaker than $\nu_e$, indicating that the transition layer has to move with positive velocity in order to satisfy the number flux across the EXZS.

To develop the discussion more quantitatively, we show spatial distributions of $P_{ee}$ at multiple time snapshots for cases with $\alpha=1, 0.5,$ and $0.1$ in Fig.\,\ref{fig:scale_plot} (from top to bottom).
In this figure, the horizontal axis ($\lambda$) corresponds to the z-axis normalized by the width between the head of $\nu_e$ beam ($\lambda = 1$) and $\bar{\nu}_e$ ($\lambda = 0$).
The $\lambda-$coordinate is useful to show results at different time snapshots in the same scale.
A noticeable feature appearing in Fig.\,\ref{fig:scale_plot} should be mentioned here.
The transition layer (or the position of EXZS, $\lambda = \lambda_{\mathrm{EXZS}}$) is nearly time-independent in the $\lambda$-space, which reflects that the layer is determined by the conservation law of neutrinos in FFC.
This is because the total $\nu_e$ that passing through $\lambda = \lambda_{\mathrm{EXZS}}$ (i.e., undergoing FFS) is $G_{v_z=1} \times (1-\lambda_{\mathrm{EXZS}})$, while that of $\bar{\nu}_e$ is $G_{v_z=-1} \times \lambda_{\mathrm{EXZS}}$.
Assuming FFS, both of them are equal to each other, leading to $\lambda_{\mathrm{EXZS}} = 1/(1 + \alpha)$.
This illustrates that the transition layer is constant in time in $\lambda$-space.
Indeed, our numerical simulations show that EXZS is located at $\lambda_{\mathrm{EXZS}} \sim 1/(1 + \alpha)$ (see dotted vertical lines in Fig.\,\ref{fig:scale_plot}).

This argument for the position of EXZS in $\lambda$-space is also useful to estimate the moving velocity of EXZS in real $z$-space.
We note that the area between the head of $\nu_e$ and $\bar{\nu}_e$ expands with time ($\Delta L_z = 2c \times \Delta t$, where $\Delta L_z$ and $c$ denote the expansion length in the time interval of $\Delta t$ and the speed of light, respectively), indicating that the EXZS position in $z$-space becomes time-dependent in general.
The velocity of EXZS ($v_{\mathrm{EXZS}}$) can be estimated as,
\begin{equation}
v_{\mathrm{EXZS}} = 2 c \times (\lambda_{\mathrm{EXZS}} - 0.5) = \frac{1-\alpha}{1+\alpha} \, c.
\end{equation}
This is a rationale behind the time evolution of EXZS displayed in Fig.\,\ref{fig:surv_time} and also those found in global FFC simulations in BNSM \cite{Nagakura:2023b}.

\begin{figure}[t]
    \centering
    \includegraphics[width=1.\linewidth]{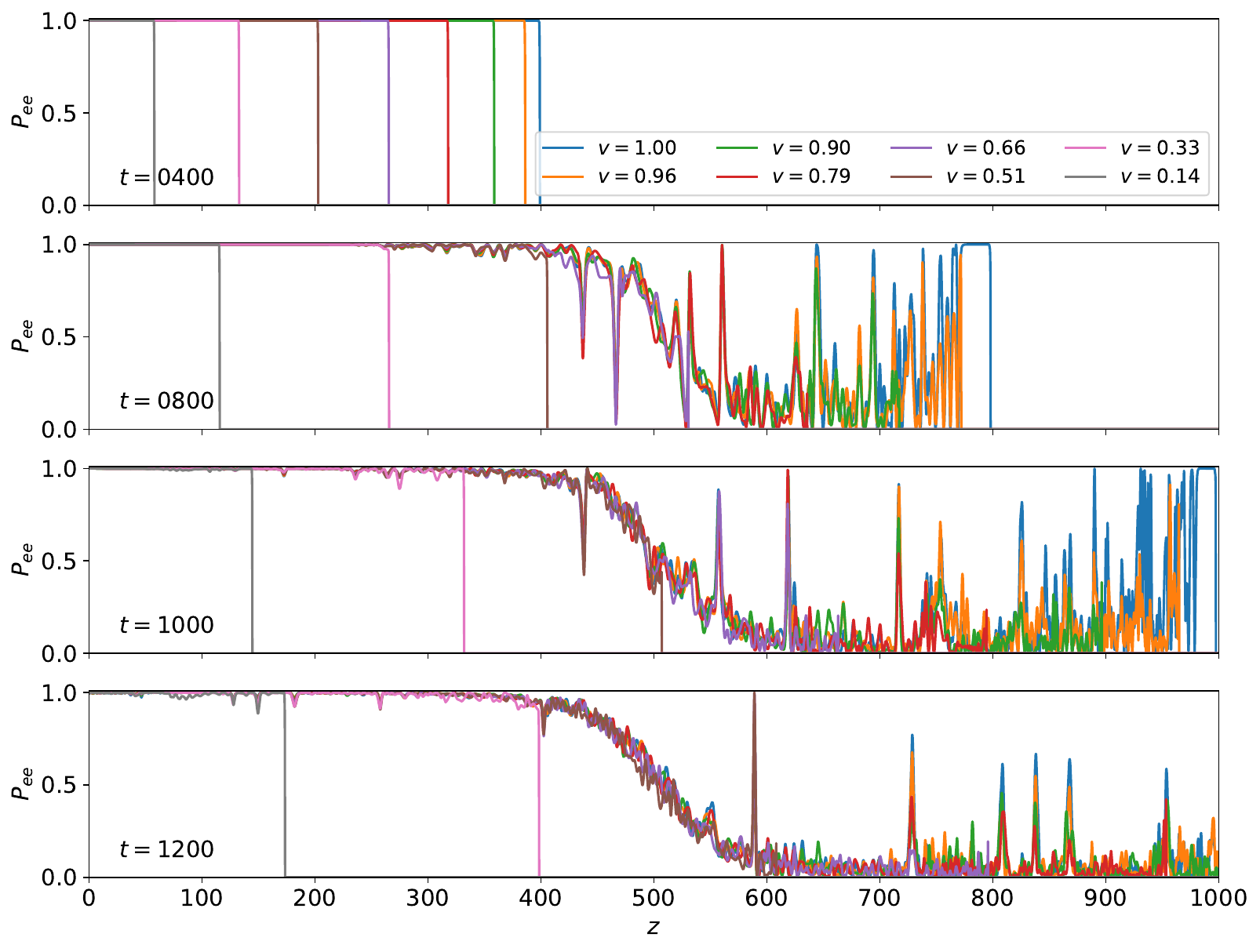}
    \caption{Time evolution of survival probability in the multiple-beams model ($N_{v_z}=32$) with $\alpha=1$.
    The positive-velocity components are only shown.
    }
    \label{fig:multiple_beam}
\end{figure}

One may wonder if FFS occurs in multiple beams.
In Fig.\,\ref{fig:multiple_beam}, we show time- and spatial distributions of $P_{ee}$ for the same simulation with $\alpha=1$ but with multiple beams ($N_{v_z}=32$).
In this simulation, we inject 16 $\nu_e$-beams and the same number of $\bar{\nu}_e$-beams from $z=0$ and $L_z$, respectively.
The number density of neutrinos for each ray is assumed to be the same, while the flight direction is distributed in $0 (-1) \le v_{z} \le 1 (0)$ for $\nu_e (\bar{\nu}_e)$ according to Gauss-Legendre quadrature.
In the early phase, we find that there are some quantitative differences from the case with the two colliding beams;
for instance, the onset time when flavor conversion appears is delayed.
This is simply because the self-interaction potential effectively becomes weaker due to the fact that relative angles between $\nu_e$ and $\bar{\nu}_e$ are narrower than the head-on collision.
On the other hand, the overall property is the same in the late phase;
all neutrinos passing through EXZS undergo FFS, and the EXZS stagnates stably at $z \sim 500$.

\section{Mechanism of flavor swap}
We now turn our attention to the physical mechanism of FFS.
We consider the mechanism of FFS with the expression of polarization vector of neutrinos: $\rho = \mathrm{Tr}\rho/2 + \boldsymbol{P}\cdot\boldsymbol{\sigma}/2$, where $\boldsymbol{\sigma}$ denotes the Pauli matrices.
The third-component (hereafter denoted as $P_3$) represents the degree of flavor eigenstate;
$P_3=1$ and $-1$ correspond to pure $\nu_e$ and $\nu_x$ states, respectively.

We first rewrite QKE for the colliding two-neutrino beam model in terms of $\boldsymbol{P}$, which can be expressed as
\begin{align}
    \partial_t \boldsymbol{P} + \partial_z \boldsymbol{P} &= -2\mu\alpha
    \bar{\boldsymbol{P}}\times \boldsymbol{P}, \label{eq:Ptwob} \\
    \partial_t \bar{\boldsymbol{P}} - \partial_z \bar{\boldsymbol{P}} &= 2\mu \boldsymbol{P}\times \bar{\boldsymbol{P}},
    \label{eq:Pbtwob} 
\end{align}
where we assume that $\boldsymbol{P}$ ($\bar{\boldsymbol{P}}$) is non-trivial only for $v_z=1 (-1)$.
It should also be mentioned that the norm of $\boldsymbol{P}$ and $\bar{\boldsymbol{P}}$ are constant in time and space where they are finite because of the absence of incoherent collision terms;
\begin{equation}
   (\partial_{t}+\partial_{z}) \lVert\boldsymbol{P}\rVert^2 = 2\boldsymbol{P}\cdot\left[-2\mu\alpha \bar{\boldsymbol{P}}\times\boldsymbol{P}\right] = 0.
   \label{eq:Pnormcon}
\end{equation}
For convenience, we perform a coordinate transformation: $t^{\prime}=t$ and $z^{\prime}=z-ct$.
This corresponds to a coordinate shifting with $v_z=1$ (we note that the coordinate bases for $t^{\prime}$ and $z^{\prime}$ are no longer orthogonal to each other).
In the coordinate, Eqs.\,\eqref{eq:Ptwob} and \eqref{eq:Pbtwob} can be transformed to
\begin{align}
   \partial_{t^{\prime}} \boldsymbol{P} &= -2\mu\alpha \bar{\boldsymbol{P}}\times \boldsymbol{P}\\
   \partial_{t^{\prime}} \bar{\boldsymbol{P}} - 2 \partial_{z^{\prime}} \bar{\boldsymbol{P}} &= 2\mu \boldsymbol{P}\times \bar{\boldsymbol{P}}.
\end{align}
Then, the second time derivative of $P_{3}$ can be written as
\begin{align}
    \partial_{t^{\prime}}^2 {P}_3 
    =  -4\mu^2\alpha & 
    \Biggl\{ \frac{1}{\mu} \Bigl[ \left(\partial_{z^{\prime}} \bar{\boldsymbol{P}}\right) \times \boldsymbol{P} \Bigr] \cdot \hat{\boldsymbol{e}}_3 \notag \\
    & + \left(\alpha P_3+\bar{P}_3\right) - \left(\boldsymbol{P}\cdot\bar{\boldsymbol{P}}\right)\left(P_3+\alpha\bar{P}_3\right) \Biggr\},
    \label{eq:acceleration}
\end{align}
where $\hat{\boldsymbol{e}}_3$ denotes the third coordinate basis in flavor space.
Hereafter, let us consider the asymptotic state ($t \to \infty$) of $P_3$.
In the colliding beam model, $\bar{\boldsymbol{P}}$ at $t \to \infty$ is given from a boundary condition:
$\bar{P}_3=1$ and $\partial_{z^{\prime}} \bar{\boldsymbol{P}}=0$ (since we inject $\bar{\nu}_e$ at the boundary).
This indicates that Eq.\,\eqref{eq:acceleration} can be approximated as,
\begin{equation}
    \partial_{t^{\prime}}^2 {P}_3 
    \sim  -4\mu^2\alpha \Bigl[ 1 - (P_3)^2 \Bigr],
    \label{eq:acceleration_v2}
\end{equation}
where we use the relation $\boldsymbol{P}\cdot\bar{\boldsymbol{P}}\sim P_3$ in the asymptotic state of $\nu_e$ where $\bar{\boldsymbol{P}}=\hat{\boldsymbol{e}}_3$.
This suggests that the asymptotic state of neutrinos satisfies $P_3=1$ or $-1$.
However, only $P_3=1$ (i.e., pure $\nu_e$ state) is clearly unstable because $\partial_{t^{\prime}}^2 P_3$ is always negative, implying that FFC occurs during neutrino evolution.
Eq.\,\eqref{eq:acceleration_v2} also shows that the FFC can not stop at $P_3=0$ (flavor equipartition), and flavor conversion proceeds until FFS is achieved ($P_3=-1$) and the system becomes stable.
It is also worth to note that the same conclusion holds for $\alpha > 0$, indicating that the flavor swap eventually takes place in the presence of $\bar{\nu}_e$ (or negative ELN-XLN).

\begin{figure}[t]
    \centering
    \includegraphics[width=1.\linewidth]{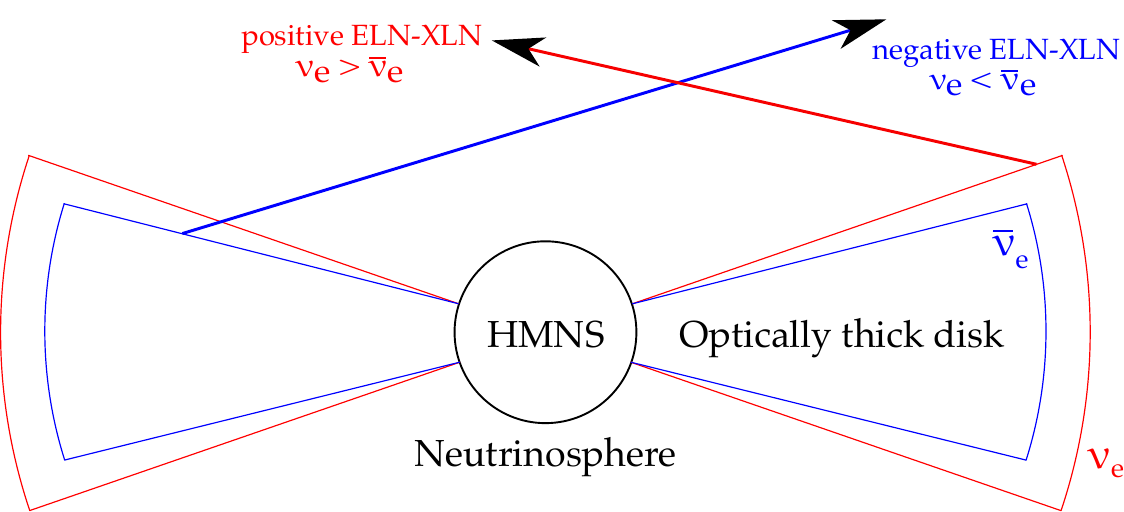}
    \caption{Schematic pictures of neutrino rays from emission surfaces in BNSM.
    Neutrinos are emitted from both hypermassive neutron star (HMNS) and the surrounding disk.
    Shell structure in the emission surface expresses flavor-dependent neutrino sphere.}
    \label{fig:schematic_vbeam}
\end{figure}

The above argument illustrates the difference from the cases where FFC achieves flavor equipartitions.
In the case with flavor equipartition, both neutrinos and antineutrinos are co-evolved, indicating that the asymptotic state of $\bar{\boldsymbol{P}}$ is determined by their interplay, leading to $\bar{P}_3 \sim 0$ at the time when $\partial_{z^{\prime}} \bar{P}_3 = 0$ \cite{Zaizen:2023,Zaizen:2023a,Xiong:2023b}.
In cases of colliding beam models, however, the neutrinos and antineutrinos evolve separately, and the asymptotic state of the other neutrinos/antineutrinos has already been given at the opposite boundary, which leads to FFS.

One thing we should mention is that BNSM remnants can naturally offer similar environments for occurrences of FFS.
As described in Ref.\,\cite{Wu:2017}, $\bar{\nu}_e$ usually has higher emission than $\nu_e$, but the $\nu_e$ can dominate over $\bar{\nu}_e$ around the outer edge of $\nu_e$ sphere.
This indicates that positive and negative ELN rays intersect with each other, leading to ELN angular crossings, as schematically illustrated in Fig.\,\ref{fig:schematic_vbeam}.
Note that XLN is zero in the neutrino radiation field without flavor conversion.
This is the similar situation in the colliding neutrino beam model, whereas it is unlikely for CCSNe to generate such globally aspherical geometry.

\section{Conclusions}
In this paper, we presented the novel dynamics of fast neutrino-flavor swap (FFS), which would appear in the geometry of BNSMs.
Since neutrinos in BNSMs are emitted from the surrounding disk, a broader angular distribution of neutrinos, including a nearly head-on collision, emerges and potentially has an angular crossing between $\nu_e$ and $\bar{\nu}_e$ with almost opposite characteristic velocities, whereas the environment is clearly in contrast to CCSN.
Our colliding neutrino beam model captures the essential feature of such a geometry; in fact, we demonstrated FFS by the simple model.
We also presented that the geometry and velocity of ELN-XLN Zero Surface (EXZS or a transition layer of flavor conversion) are determined so as to satisfy the conservation of the ELN-XLN number flux across the EXZS.
In the domain normalized by the width between the heads of two neutrino beams, the transition layer is stationary irrespective of the flavor asymmetry while its location is given by it.
This observation in our numerical results provides a rationale behind the time-dependent feature of EXZS in Ref.\,\cite{Nagakura:2023b}.

We also analytically showed the mechanism of FFS based on polarization vectors for neutrino density matrices.
Considering the asymptotic state of $\nu_e$, we find that FFC does not stop until the flavor swap completes.
This is because $\bar{\nu}_e$ (or negative ELN-XLN), propagating in the opposite direction to $\nu_e$, continues to be supplied from the opposite boundary.
This exhibits that the occurrence of FFS hinges on the ELN-XLN distributions and the geometry of neutrino spheres, which are clearly different from the cases where FFC evolves towards flavor equipartition.

Since FFS corresponds to the most extreme case of flavor conversion, it would yield a significant change in the neutrino radiation field, particularly neutrino absorption. 
This affects the electron fraction of ejecta and subsequently r-process nucleosynthesis and kilonova.
In CCSNe, FFS would not occur and neutrinos would achieve a flavor equipartition through normal FFC, but there is a caveat.
They might occur in CCSNe if large-scale coherent asymmetric neutrino emission appears in the CCSN core.
It would be realized when the CCSN core is rapidly rotating, strong lepton-number emission of self-sustained asymmetry (LESA) occurs \cite{Tamborra:2014a,Glas:2019,Powell:2019}, or the proto-neutron star is accelerated by asymmetric neutrino emission \cite{Nagakura:2019b}.
Addressing issues whether FFS can emerge in such extreme CCSN environments requires a more detailed investigation, which is deferred to our future work.
\\

\begin{acknowledgments}
We would like to express special thanks to Manu George for sharing his unpublished numerical results which inspired our colliding neutrino beam model.
We are also grateful to Meng-Ru Wu, Zewei Xiong, and Lucas Johns for useful comments and discussions.
We thank the Focus workshop on collective oscillations and chiral transport of neutrinos at the Academia Sinica for eliciting the discussions that led to the completion of this paper.
M.Z. is supported by the Japan Society for Promotion of Science (JSPS) Grant-in-Aid for JSPS Fellows (Grants No. 22KJ2906) from the Ministry of Education, Culture, Sports, Science, and Technology (MEXT) in Japan. H.N. is supported by Grant-inAid for
Scientific Research (23K03468).
The numerical computations were carried out on Cray XC50 at the Center for Computational Astrophysics, National Astronomical Observatory of Japan.
This work is also supported by HPCI System Research
Project (Project ID: 230033, 230204, 230270).
\end{acknowledgments}



\bibliography{papers}

\end{document}